\documentclass[sigplan,screen]{acmart}

\usepackage[frozencache,cachedir=.]{minted}

\AtBeginDocument{%
  \providecommand\BibTeX{{%
    \normalfont B\kern-0.5em{\scshape i\kern-0.25em b}\kern-0.8em\TeX}}}

\setcopyright{acmcopyright}
\copyrightyear{2022}
\acmYear{2022}
\acmDOI{XXXXXXX.XXXXXXX}

\acmConference[Conference acronym 'XX]{Make sure to enter the correct
  conference title from your rights confirmation emai}{June 03--05,
  2018}{Woodstock, NY}
\acmPrice{15.00}
\acmISBN{978-1-4503-XXXX-X/18/06}

\begin{document}

\title{Automatic compile-time synthesis of entropy-optimal Boltzmann samplers}

\author{Maciej Bendkowski}
\email{maciej.bendkowski@gmail.com}
\orcid{0000-0003-4383-4606}

\begin{abstract}
  We present a framework for the automatic compilation of multi-parametric
Boltzmann samplers for algebraic data types in Haskell. Our framework uses
Template Haskell to synthesise efficient, entropy-optimal samplers generating
random instances of user-declared algebraic data types. Users can control the
outcome distribution through a pure, declarative interface.  For instance, users
can control the mean size and constructor frequencies of generated objects. We
illustrate the effectiveness of our framework through a prototype
\texttt{generic-boltzmann-brain} library showing that it is possible to control
thousands of different parameters in systems of tens of thousands of ADTs. Our
prototype framework synthesises Boltzmann samplers capable of rapidly generating
random objects of sizes in the millions.
\end{abstract}



\maketitle

\section{Introduction}
Consider the following example of a pair of algebraic data types
\mintinline[fontsize=\small]{haskell}{Lambda} and
\mintinline[fontsize=\small]{haskell}{DeBruijn} defining lambda terms in
DeBruijn notation~\cite{deBruijn1972}:
\begin{minted}[fontsize=\small]{haskell}
data DeBruijn
  = Z
  | S DeBruijn

data Lambda
  = Index DeBruijn
  | App Lambda Lambda
  | Abs Lambda
\end{minted}
In the following paper we develop a general framework for
compile-time generation of efficient Boltzmann samplers~\cite{DuFlLoSc} for system of algebraic data
types, such as~\mintinline[fontsize=\small]{haskell}{Lambda} and
\mintinline[fontsize=\small]{haskell}{DeBruijn}.
Our prototype
library\footnote{\url{https://github.com/maciej-bendkowski/generic-boltzmann-brain}}
exposes a minimal, declarative Template Haskell~\cite{template-haskell}
interface. For instance
\begin{minted}[fontsize=\small]{haskell}
  mkDefBoltzmannSampler ''Lambda 10_000
\end{minted}
declares \mintinline{haskell}{Lambda} an instance of the
\mintinline{haskell}{BoltzmannSampler} type class:
\begin{minted}[fontsize=\small]{haskell}
class BoltzmannSampler a where
  sample :: RandomGen g =>
            UpperBound ->
            MaybeT (BuffonMachine g) (a, Int)
\end{minted}
The above type class defines types with a single
\mintinline[fontsize=\small]{haskell}{sample} function.  Given an integer upper
bound $n$, \mintinline[fontsize=\small]{haskell}{sample} generates a random
instance $\gamma$ of type \mintinline[fontsize=\small]{haskell}{a} together with its
corresponding size $s \leq n$. While computing a random
\mintinline[fontsize=\small]{haskell}{a}, the generator consumes random bits
provided within a custom \mintinline[fontsize=\small]{haskell}{BuffonMachine g}
monad. Because the generation process might sometimes fail, the whole
computation is wrapped in a \mintinline[fontsize=\small]{haskell}{MaybeT} monad
transformer.

The \mintinline[fontsize=\small]{haskell}{sample} function satisfies two key
Boltzmann sampler properties:
\begin{itemize}
  \item instances of \mintinline[fontsize=\small]{haskell}{a} with the \emph{exact
same size} have \emph{the exact same} probability of being generated, and
  \item the \emph{expected size} of the generated instances of
        \mintinline[fontsize=\small]{haskell}{a} follows the user-declared value,
        such as $10,000$ for \mintinline[fontsize=\small]{haskell}{Lambda}.
\end{itemize}
In other words, while the size of the outcomes may vary, the outcome
distribution is \emph{fair},~\emph{i.e.} \emph{uniform} when conditioned on the
size of the generated objects.

When a finer control over the outcome size is required, \emph{rejection sampling}
can be adopted~\emph{cf.}~\cite{BodLumRolin}:
\begin{minted}[fontsize=\small]{haskell}
rejectionSampler ::
  (RandomGen g, BoltzmannSampler a) =>
      LowerBound -> UpperBound -> BuffonMachine g a
\end{minted}
Given two lower and upper bounds, a rejection sampler generates random instances
of \mintinline[fontsize=\small]{haskell}{a} until a sample of admissible size is
generated. The expected runtime complexity of such a sampler depends on the
\emph{width} of the admissible size window. If it is an interval of the form
$[(1-\varepsilon)n, (1+\varepsilon)n]$ for some positive \emph{tolerance} parameter $\varepsilon > 0$, the
runtime complexity of the rejection sampler is
\emph{linear},~\emph{i.e.}~$O(n)$. When the tolerance parameter $\varepsilon$ is equal to
$0$, the rejection sampler returns objects of some constant size $n$, and the
expected runtime of \mintinline[fontsize=\small]{haskell}{rejectionSampler}
becomes $O(n^{2})$,~\emph{cf.}~\cite{BodLumRolin,DuFlLoSc}.

Compiled rejection samplers are readily available for use in property testing
frameworks, such as QuickCheck~\cite{Claessen-2000}. For instance,
\mintinline[fontsize=\small]{haskell}{BuffonMachine g} computations can be easily
converted to QuickCheck's \mintinline[fontsize=\small]{haskell}{Gen} values:
\begin{minted}[fontsize=\small]{haskell}
quickCheckRejectionSampler ::
  BoltzmannSampler a =>
    (Int -> (LowerBound, UpperBound)) -> Gen a
\end{minted}

By default, the \emph{size} of generated objects is equal to the overall
\emph{weight} of constructors used in their construction. For instance, the size
of \mintinline[fontsize=\small]{haskell}{Abs (App (Index Z) (Index Z))} is equal
to six as it consists of size constructor of default weight one. If such a
\emph{size notion} is not desired, it is possible to redefine the constructor
weights, \emph{e.g.} as follows:

\begin{minted}[fontsize=\small]{haskell}
mkBoltzmannSampler
  System
    { targetType = ''Lambda
    , meanSize = 10_000
    , frequencies = def
    , weights =
        ('Index, 0)
          <:> $(mkDefWeights ''Lambda)
    }
\end{minted}
Note that here we declared a Boltzmann sampler for \mintinline[fontsize=\small]{haskell}{Lambda}
with (expected) mean size $10,000$, and a new set of \emph{constructor weights} in which
all constructors except \mintinline[fontsize=\small]{haskell}{Index} have default weight one.
The remaining \mintinline[fontsize=\small]{haskell}{Index} constructor contributes now weight
zero to the overall size of lambda terms.

\subsection{Beyond uniform outcome distribution}
In~\cite{BodPonty} a generalisation of Boltzmann samplers was introduced which
lifted the classic univariate Boltzmann samplers to a \emph{multi-parametric} setting.
This multivariate paradigm is reflected in the presented framework in form of
custom \emph{constructor frequencies}. For instance

\begin{minted}[fontsize=\small]{haskell}
mkBoltzmannSampler
  System
    { targetType = ''Lambda
    , meanSize = 10_000
    , frequencies = ('Abs, 4_000) <:> def
    , weights =
        ('Index, 0)
          <:> $(mkDefWeights ''Lambda)
    }
\end{minted}
declares a multi-parametric Boltzmann sampler for \mintinline[fontsize=\small]{haskell}{Lambda}
in which the target mean size is still $10,000$, however now we additionally require
that the \emph{mean weight contribution} of abstractions is equal to $4,000$.

The \mintinline[fontsize=\small]{haskell}{size} function satisfies now
the following generalised Boltzmann sampler properties:
\begin{itemize}
  \item instances of \mintinline[fontsize=\small]{haskell}{Lambda} with the exact
        same size \emph{and} the same cummulative abstraction weights have \emph{the exact same}
        probability of being generated, and
  \item the \emph{expected size} of the generated instances is still
        $10,000$, whereas the \emph{expected number} of abstractions is
        equal to the user-declared value of $4,000$.
\end{itemize}

It is therefore possible to \emph{tune} the natural frequency of each
constructor in \mintinline[fontsize=\small]{haskell}{Lambda} and
\mintinline[fontsize=\small]{haskell}{DeBruijn} to one's needs. Note however
that such an additional control causes a significant change in the underlying
outcome distribution. In extreme cases, such as for instance requiring 80\% of
internal nodes in plane binary trees, the sampler might fail to compile or be
virtually ineffective due to the sparsity of tuned structures.

\subsection{Multiple Boltzmann sampler instances}
Because Boltzmann samplers are implemented as instances of the
\mintinline[fontsize=\small]{haskell}{BoltzmannSampler} type class, we cannot
have two distinct Boltzmann samplers for the same type \mintinline[fontsize=\small]{haskell}{a}.
In some circumstances, however, having multiple Boltzmann samplers with
different constructor frequencies or even size notions might be beneficial. To enable such use cases,
the presented framework lets users define Boltzmann samplers for
\mintinline[fontsize=\small]{haskell}{newtype}s
of respective types.

For instance, in the following snippet we define a representation of so-called
\emph{binary lambda terms}, initially introduced by Tromp~\cite{tromp} for
the purpose of using lambda calculus in algorithmic information theory (\emph{cf.} also~\cite{grygiel_lescanne_2015}):

\begin{minted}[fontsize=\small]{haskell}
newtype BinLambda = MkBinLambda Lambda

mkBoltzmannSampler
  System
    { targetType = ''BinLambda
    , meanSize = 12_000
    , frequencies = ('Abs, 3000) <:> def
    , weights =
        ('Index, 0)
          <:> ('App, 2)
          <:> ('Abs, 2)
          <:> $(mkDefWeights ''Lambda)
    }
\end{minted}
The \mintinline[fontsize=\small]{haskell}{BinLambda} type borrows the algebraic representation
of \mintinline[fontsize=\small]{haskell}{Lambda}.  Custom weights for \mintinline[fontsize=\small]{haskell}{App}
and \mintinline[fontsize=\small]{haskell}{Abs} reflect Tromp's recursive binary string representation of lambda terms:
\begin{minted}[fontsize=\small]{haskell}
  encode :: Lambda -> [Bool]
  encode = \case
    Abs t     -> False : False : encode t
    App lt rt -> False : True : encode lt ++ encode rt
    Index n   -> encode' n
  where
    encode' :: DeBruijn -> [Bool]
    encode' = \case
      S n' -> True : encode' n'
      Z    -> [True]
\end{minted}
Note that the size of a binary lambda term corresponds to the length of the corresponding
encoded binary string.
In addition to a new size notion, \mintinline[fontsize=\small]{haskell}{BinLambda}
uses a different set of constructor frequencies, and mean size.

\section{Univariate Boltzmann models}
Before explaining the general architecture of our framework let us pause for a
moment and focus on the mathematical foundations of Boltzmann samplers,~\emph{cf.}~\cite{DuFlLoSc}.

Let $\mathcal{S}$ be a set of objects endowed with an intrinsic \emph{size}
function $|\cdot| \colon \mathcal{S} \to \mathbb{N}$ with the property that that for
all $n \in \mathbb{N}$, the set of objects of $n$ in $\mathcal{S}$ is finite.  For
such a class of objects, the corresponding (univariate) \emph{generating
function} $S(z)$ is the power series $S(z)$ defined as
\begin{equation}
  S(z) = \sum_{n \geq 0} s_{n} z^{n}
\end{equation}
whose coefficients $\left(s_{n}\right)_{n \geq 0}$ denote the number of objects of
size $n$ in $\mathcal{S}$, \emph{cf.}~\cite{wilf}.

Given a real \emph{control parameter} $x \in [0, 1]$, a \emph{Boltzmann model}~\cite{DuFlLoSc} is a
probability distribution in which the probability $\mathbb{P}_{x}(\omega)$ of
generating an object $\omega \in \mathcal{S}$ satisfies
\begin{equation}
  \mathbb{P}_{x}(\omega) = \frac{x^{|\omega|}}{S(x)}.
\end{equation}
provided that $S(x)$ is finite\footnote{Generating functions corresponding to
algebraic specifications discussed in the current paper are \emph{analytic},
\emph{i.e.}~are convergent within some non-empty complex circle $|z| < \rho$ for
$\rho \in \mathbb{R}$ depending only on the class $\mathcal{S}$.}.

\begin{figure}
\centering
\includegraphics[width=\columnwidth]{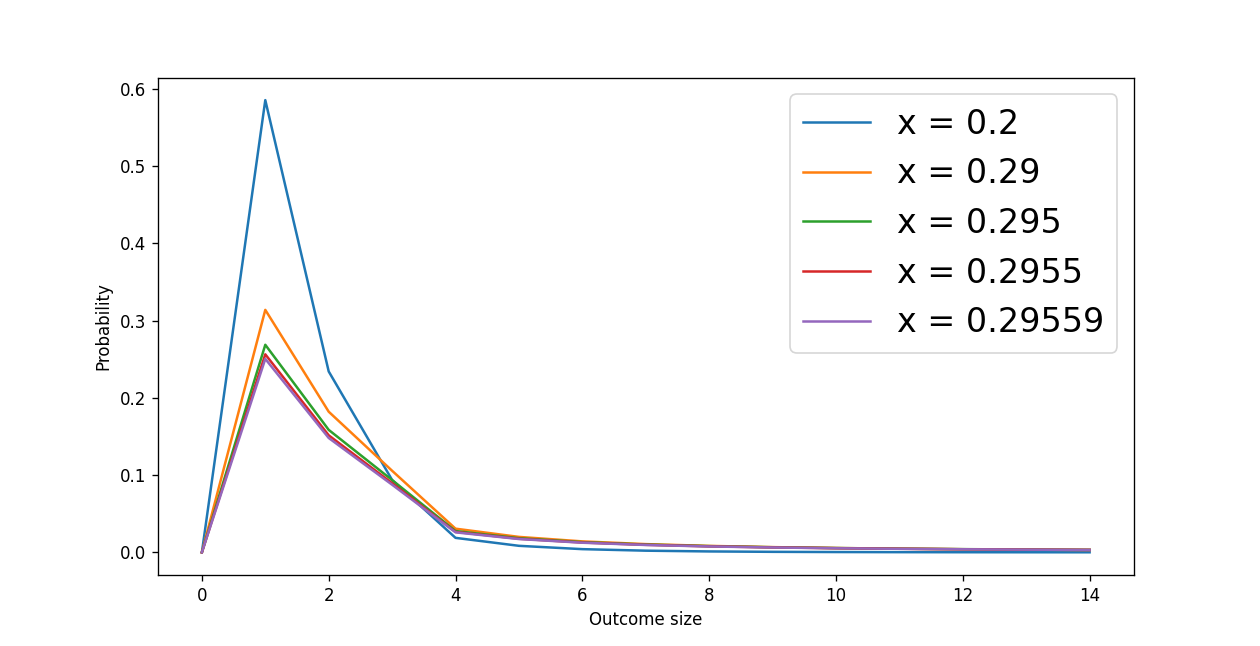}
\caption{Example univariate Boltzmann models for \mintinline[fontsize=\small]{haskell}{Lambda}.
  Note that the values $\mathbb{P}_{x}(N = n)$ quickly approach zero yet never reach it.}\label{fig:boltzmann-model-lambda}
\end{figure}

Note that under such a model
\begin{itemize}
  \item objects of equal size have equal probabilities, and
  \item the outcome size is varying random variable.
\end{itemize}
Indeed, note that the probability $\mathbb{P}_{x}(N = n)$ that
the size $N$ of a randomly generated object is equal to $n$ satisfies
\begin{equation}
  \mathbb{P}_{x}(N = n) = \frac{s_{n} x^{n}}{S(x)}
\end{equation}
In other words, the outcome size distribution depends both on
the control parameter $x$, as well as on the intrinsic size distribution in
$\mathcal{S}$, see~\emph{e.g.}~\autoref{fig:boltzmann-model-lambda}.

In consequence, the control parameter $x$ influences the \emph{expected (mean)
outcome size} $\mathbb{E}_{x}(N)$, as well as the standard deviation $\sigma_{x}(N)$:
\begin{equation}\label{eq:boltzmann-model-expected-size}
\mathbb{E}_{x}(N) = x \frac{\frac{d}{d x}S(x)}{S(x)}
\quad \sigma_{x}(N) = \sqrt{x \frac{d}{d x} \mathbb{E}_{x}(N)}
\end{equation}
Given access to the values of $S$ and its derivative $\frac{d}{d x} S$ it is
possible to use formula~\eqref{eq:boltzmann-model-expected-size} and aptly
choose a value of the control parameter $x$ so to obtain a Boltzmann model with
expect size of outcomes equal to a \emph{target mean size} $n$. Even though, in
general, explicit formulas or numerical oracles for $S(x)$ and
$\frac{d}{d x}S(x)$ might not be readily available, we will soon see that for
specifications corresponding to algebraic data types, we can construct efficient
oracles and thus automatically find apt values for the control parameter.

\subsection{Compiling Boltzmann samplers}
Boltzmann samplers, realising the outcome Boltzmann model, follow closely the
\emph{sum-of-products} structure of ADTs and hence can be compiled in a
recursive fashion.

\subsubsection{Singletons}
Consider a singleton class $\mathcal{S}$, \emph{i.e.} a set consisting of a
single element $\gamma$.  Note that the corresponding generating function takes the
form $S(z) = z^{|\gamma|}$. Consequently, the probability $\mathbb{P}_{x}(\gamma)$ of
sampling $\gamma$ is equal to one, and the respective Boltzmann sampler always returns
$\gamma$.

\subsubsection{Products}
Consider a product class $\mathcal{S}$ consisting of \emph{pairs} $\gamma = (\alpha, \beta)$ where the
components are arbitrary elements of classes $\mathcal{A}$ and $\mathcal{B}$, and $|\gamma| = |\alpha| + |\beta|$.
Under a Boltzmann model the probability $\mathbb{P}_{x}(\gamma)$ that $\gamma$ is sampled satisfies
\begin{equation}\label{eq:boltzmann-model-pair-i}
  \mathbb{P}_{x}(\gamma) = \frac{x^{|\gamma|}}{S(x)}
\end{equation}
Since $|\gamma| = |\alpha| + |\beta|$ we can rewrite
rewrite~\eqref{eq:boltzmann-model-pair-i} as
\begin{equation}\label{eq:boltzmann-model-pair-ii}
  \mathbb{P}_{x}(\gamma) = \frac{x^{|\gamma|}}{S(x)} = \frac{x^{|\alpha| + |\beta|}}{S(x)}
  = \frac{x^{|\alpha|} x^{|\beta|}}{S(x)}
\end{equation}
Now, let us notice that $S(z) = A(z) B(z)$ as
\begin{align}
  \begin{split}
  \left(\sum_{n \geq 0} a_n z^n\right) \cdot \left(\sum_{n \geq 0} b_n z^n\right) &= \sum_{n \geq 0} c_n z^n \\
  \text{where} \quad c_n &=\sum_{k=0}^n a_k b_{n-k}
  \end{split}
\end{align}
following Cauchy's product formula for power series.
Indeed, the number of pairs $\mathcal{A} \times \mathcal{B}$ of size $n$ is equal
to $\sum_{k=0}^n a_k b_{n-k}$ where $\{a,b\}_i$ denotes the number
of objects in $\mathcal{A}$ (respectively $\mathcal{B}$) of size $i$. Therefore

\begin{equation}\label{eq:boltzmann-model-pair-iii}
  \mathbb{P}_{x}(\gamma)
  = \frac{x^{|\alpha|} x^{|\beta|}}{S(x)}
  = \frac{x^{|\alpha|} x^{|\beta|}}{A(x) B(x)}
  = \mathbb{P}_{x}(\alpha) \mathbb{P}_{x}(\beta)
\end{equation}
It means that in order to generate a random pair $\gamma$ corresponding to $\mathcal{S}$ using
a Boltzmann sampler, we can invoke Boltzmann samplers for $\mathcal{A}$ and $\mathcal{B}$ using the
same control parameter $x$, and then return a pair of their results. Note that the
same principle naturally generalises onto for tuples of arbitrary length as
  $(a, b, c) \cong ((a, b), c) \cong (a, (b, c))$.

\subsubsection{Coproducts}
Consider a coproduct class $\mathcal{S} = \mathcal{A} + \mathcal{B}$ which is a \emph{disjoint sum} of two
classes $\mathcal{A}$ and $\mathcal{B}$. In other words, $\mathcal{S}$ consists
of elements $\gamma$ which belong to either $\mathcal{A}$ or $\mathcal{B}$, but not
both at the same time. Note that in such a case the probability
$\mathbb{P}_{x}(\gamma \in \mathcal{A})$ that an arbitrary object $\gamma$ in $\mathcal{A}$
is sampled satisfies
\begin{equation}
  \mathbb{P}_{x}(\gamma \in \mathcal{A}) = \frac{A(x)}{S(x)} \quad \text{as} \quad
  A(x) = \sum_{\gamma \in \mathcal{A}} x^{|\gamma|}
\end{equation}
It means that in order to generate a random object $\gamma$ in $\mathcal{S}$ using a
Boltzmann sampler, we have to make a \emph{skewed} coin toss. With probability
$\frac{A(x)}{S(x)}$ we invoke the sampler corresponding to $\mathcal{A}$, and
with probability $\frac{B(x)}{S(x)}$ we invoke the sampler corresponding to
$\mathcal{B}$.
Like in the case of products, the same principle
naturally generalises onto arbitrary sums as
  $a + b + c \cong (a + b) + c \cong a + (b + c)$.

\subsubsection{Algebraic data types}
The above simple Boltzmann sampler compilation rules can be readily applied to concrete
algebraic data types. Consider our running example system of two ADTs
\mintinline[fontsize=\small]{haskell}{Lambda} and
\mintinline[fontsize=\small]{haskell}{DeBruijn}.

A Boltzmann sampler for \mintinline[fontsize=\small]{haskell}{Lambda} has to first make
a random decision which constructor to use,~\emph{i.e.} \mintinline[fontsize=\small]{haskell}{Abs},
\mintinline[fontsize=\small]{haskell}{App}, or \mintinline[fontsize=\small]{haskell}{Index}.
This decision follows the co-product compilation rule.

If \mintinline[fontsize=\small]{haskell}{Abs} is chosen, following the product rule, the \mintinline[fontsize=\small]{haskell}{Lambda}
Boltzmann sampler has to invoke a Boltzmann sampler for \mintinline[fontsize=\small]{haskell}{Lambda} (\emph{i.e.}~itself), generate a random lambda term \mintinline[fontsize=\small]{haskell}{lt},
and output \mintinline[fontsize=\small]{haskell}{Abs lt}. Likewise, if \mintinline[fontsize=\small]{haskell}{App}
is chosen, the \mintinline[fontsize=\small]{haskell}{Lambda} Boltzmann sampler has to
invoke itself twice, generating two random lambda terms \mintinline[fontsize=\small]{haskell}{lt}
and \mintinline[fontsize=\small]{haskell}{lt'}, and output \mintinline[fontsize=\small]{haskell}{App lt lt'}.
Finally, if \mintinline[fontsize=\small]{haskell}{Index} is chosen, the \mintinline[fontsize=\small]{haskell}{Lambda}
Boltzmann sampler has to invoke the Boltzmann sampler for \mintinline[fontsize=\small]{haskell}{DeBruijn} which will
return a random \mintinline[fontsize=\small]{haskell}{DeBruijn} index, and wrap it around \mintinline[fontsize=\small]{haskell}{Index}.
The Boltzmann sampler for \mintinline[fontsize=\small]{haskell}{DeBruijn} is constructed similarly.

Let us remark that while Boltzmann samplers readily apply to algebraic data types, they are not limited to
them. Over the years Boltzmann samplers have enjoyed a series
of extensions and improvements including, \emph{inter alia}, the support for
so-called labelled~\cite{DuFlLoSc}, Pólya~\cite{flajolet2007boltzmann}, or
first-order differential specifications~\cite{BODINI20122563}.

\section{Multivariate Boltzmann models}
The classical, univariate Boltzmann model controls a single system parameter,
\emph{i.e.} the expected outcome size. In some circumstances, however, a finer
control over the outcome distribution is required. Multivariate Boltzmann
models, initially introduced in~\cite{BodPonty}, address this issue by
generalising classical Boltzmann models to a multivariate setting in which
multiple outcome parameters can be controlled simultaneously\footnote{Let us
remark that, unless NP = RP, controlling the \emph{exact} values of multiple
parameters is practically infeasible, see~\cite{bendkowski_bodini_dovgal_2021}.}.

Analogously to their univariate counterparts, multiparametric Boltzmann models depend on
\emph{multivariate generating functions}.
A multivariate \emph{generating function} $S(z_{1},\ldots,z_{d})$ is a power series
$S(z_{1},\ldots,z_{d})$ defined as
\begin{equation}
  S(z_{1},\ldots,z_{d}) = \sum_{n_{1},\ldots,n_{d} \geq 0} s_{n_{1},\ldots,n_{d}} \prod_{i=1}^{d} z_{i}^{n_{i}}
\end{equation}
whose coefficients $\left(s_{n_{1},\ldots,n_{d}}\right)_{n \geq 0}$ denote the number of objects with
$n_{i}$ atoms of type $z_{i}$ in $\mathcal{S}$, \emph{cf.}~\cite{flajolet09}. For instance,
$z_{1}$ can correspond to the \emph{size} of lambda terms in \mintinline[fontsize=\small]{haskell}{Lambda},
whereas $z_{2}$ can denote the number of its abstractions. Then, the coefficient $s_{n,k}$ denotes the number
of lambda terms of size $n$ which have $k$ abstractions in total.

Given a vector of real control parameters $\vec{x} = \left(x_{1},\ldots,x_{d}\right)$, a \emph{multivariate Boltzmann model}
is a probability distribution in which the probability $\mathbb{P}_{\vec{x}}(\omega)$ of generating
an object $\omega \in \mathcal{S}$ with $n_{i}$ atoms of type $z_{i}$ satisfies
\begin{equation}
  \mathbb{P}_{\vec{x}}(\omega) = \frac{x_{1}^{n_{1}} \cdots x_{d}^{n_{d}}}{S(\vec{x})}.
\end{equation}
The expected number $\mathbb{E}_{\vec{x}}(N_{i})$ of atoms of type $n_{i}$ satisfies
\begin{equation}\label{eq:multivariate-boltzmann-model-expected-size}
\mathbb{E}_{\vec{x}}(N_{i}) = x_{i} \frac{\frac{\partial}{\partial_{x_{i}}}S(\vec{x})}{S(\vec{x})}
\end{equation}
Note that this is a straightforward generalisation of~\eqref{eq:boltzmann-model-expected-size}.

While compilation rules for univariate Boltzmann samplers readily generalise
onto multiparametric samplers,
\emph{cf.}~\cite{BodPonty,bendkowski_bodini_dovgal_2021}, finding apt values for
the $d$-dimensional \emph{control vector} $\vec{x}$ poses an even more
challenging problem.

\section{Parameter tuning}
The \emph{key} to compiling Boltzmann samplers with expected outcome parameters
lies in finding the value of the corresponding control vector $\vec{x}$ and
the values of respective generating functions at $\vec{x}$. We call this process
\emph{parameter tuning}.

In simple systems, such as in our single-parameter running example of
\mintinline[fontsize=\small]{haskell}{Lambda} and
\mintinline[fontsize=\small]{haskell}{DeBruijn}, we have access to analytic
closed form expressions for all the generating functions. Using the so-called
\emph{symbolic method}~\cite{flajolet09} we can lift the algebraic type
definitions onto the level generating functions corresponding to the intrinsic
size of objects in the associated classes.

Unfortunately, for most systems of algebraic data types we
do not have access to closed form expressions of respective generating functions.
For instance, the following data type
\begin{minted}[fontsize=\small]{haskell}
data T = Empty | Node T T T T T
\end{minted}
gives rise to a generating function $T(z) = z + z {T(z)}^5$ which, by the
Abel–Ruffini theorem, has no explicit closed-form form solutions.
Therefore, in general, we have to resort to \emph{numerical solutions}, instead.

For systems without additional tuning parameters we could use a quickly
convergent Newton iteration procedure developed in~\cite{pivoteau2012}. For
generalised systems with $d$ tuning parameters, on the other hand, we could
use a generalised  Newton iteration scheme developed in~\cite{BodPonty}.
Unfortunately, the latter is impractical both due to its exponential
$O(n^{{1+\frac{d}{2}}})$ running time, as well as the fact that the iteration is
convergent in an \emph{a priori} unknown $d$-dimensional vicinity of the target
control vector $\vec{x}$ value.

Given these limitations, in the actual implementation of the presented framework
we resort to an alternative method based on convex optimisation techniques.

\subsection{Convex optimisation}\label{sec:convex-optimisation}
We illustrate the principle of \emph{tuning as convex
optimisation}~\cite{bendkowski_bodini_dovgal_2021} on our running example of
\mintinline[fontsize=\small]{haskell}{Lambda} and
\mintinline[fontsize=\small]{haskell}{DeBruijn} where we request a Boltzmann
model for lambda terms with mean size $10,000$ and $2,500$ abstractions in
expectation.  We assume a size notion in which the constructor
\mintinline[fontsize=\small]{haskell}{Index} contributes weight zero and all
other constructors contribute weight one.
Let us recall the system under consideration:
\begin{minted}[fontsize=\small]{haskell}
data DeBruijn
  = Z
  | S DeBruijn

data Lambda
  = Index DeBruijn
  | App Lambda Lambda
  | Abs Lambda
\end{minted}
Let us denote the (univariate) generating function corresponding to
\mintinline[fontsize=\small]{haskell}{Lambda} and
\mintinline[fontsize=\small]{haskell}{DeBruijn} by $L(z)$ and $D(z)$,
respectively.  Based on~\eqref{eq:multivariate-boltzmann-model-expected-size} we can formulate the
following~\emph{optimisation problem}:
\begin{align}\label{eq:minimisation-problem}
  \text{Minimise} &\\
&\left(z \frac{\frac{\partial}{\partial_{z}}L(z,u)}{L(z,u)} - 10,000\right) + \left(u \frac{\frac{\partial}{\partial_{u}}L(z,u)}{L(z,u)} - 2,500\right) \nonumber\\
  \text{for } z, u. \nonumber
\end{align}
In other words, we ask for $z, u$ which result in a Boltzmann model in which
the expected size of lambda terms is $10,000$ and the mean number of abstractions
is equal to $2,500$.

Unfortunately, in such a form the optimisation
problem~\eqref{eq:minimisation-problem} is too general to use an optimisation
solver. Following~\cite{bendkowski_bodini_dovgal_2021} we therefore reformulate
it as a \emph{convex optimisation problem} exploiting the regular structure of
algebraic data types \mintinline[fontsize=\small]{haskell}{Lambda} and
\mintinline[fontsize=\small]{haskell}{DeBruijn}.

We start with mapping the input system to a system of corresponding (univariate)
generating functions
using the symbolic method~\cite{flajolet09}:
\begin{align}
D(z) &= z + z D(z) \nonumber \\
L(z) &= D(z) + z {L(z)}^{2} + z L(z)
\end{align}
The transformation is purely mechanical and follows the \emph{sum-of-products}
structure of involved algebraic type definitions.

Let us start with \mintinline[fontsize=\small]{haskell}{DeBruijn}. It has two constructors
which generate \emph{distinct} inhabitants of \mintinline[fontsize=\small]{haskell}{DeBruijn}.
We can therefore think of \mintinline[fontsize=\small]{haskell}{DeBruijn} as a disjoint sum
of two classes of objects, \emph{i.e.} the singleton class \mintinline[fontsize=\small]{haskell}{Z},
and the class \mintinline[fontsize=\small]{haskell}{S DeBruijn} of successors. The former
class has a single inhabitant of size one, hence its generating function is just $z$. The
latter class, on the other hand, consists of DeBruijn indices in the form of
\mintinline[fontsize=\small]{haskell}{S n} where \mintinline[fontsize=\small]{haskell}{n}
is itself a DeBruijn index. The topmost constructor \mintinline[fontsize=\small]{haskell}{S} contributes
weight one to each of the indexes, and so the corresponding generating function takes form
$z D(z)$ where $D(z)$ is the generating function for DeBruijn indices.

Next, let us consider \mintinline[fontsize=\small]{haskell}{Lambda}. Its type
definition consists of three constructors which give rise to three distinct
classes, \emph{i.e.} indices, applications, and abstractions. Because
\mintinline[fontsize=\small]{haskell}{Index} contributes no weight, the
respective generating function is $D(z)$.  On the other hand,
\mintinline[fontsize=\small]{haskell}{App} and
\mintinline[fontsize=\small]{haskell}{Abs} contribute weight one, and so the
corresponding generating functions for applications and abstractions take forms
$z {L(z)}^{2}$ and $z L(z)$, respectively. Note that the exponent of $L(z)$
corresponds to the arity of the respective constructor.  In general, each
constructor definition
\mintinline[escapeinside=||,mathescape=true,fontsize=\small]{haskell}{T
  |$a_{1} \ldots a_{k}$| } can be thought of as a generalised product
\mintinline[escapeinside=||,mathescape=true,fontsize=\small]{haskell}{|($\cdots$(T $a_{1}) \ldots a_{k})$| }.
Consequently, the
corresponding generating function is of form $z^{w} A_{1}(z) \cdots A_{k}(z)$ where
$w$ is the weight of \mintinline[fontsize=\small]{haskell}{T}, and
$A_{1}(z),\ldots,A_{k}(z)$ are the generating functions corresponding to the
respective argument types.

Next, for each custom constructor frequency we create a new \emph{marking variable}
and place it in the definition of the respective generating function:
\begin{align}\label{eq:example-sys}
D(z,u) &= z + z D(z,u) \nonumber \\
L(z,u) &= D(z, u) + z {L(z, u)}^{2} + z u L(z, u)
\end{align}
Note that $u$ marks now occurrences of abstractions. In other words, the coefficient $l_{n,k}$
standing by $z^{n}u^{k}$ in the generating function $L(z,u)$ denotes the number of lambda
terms of size $n$ and $k$ abstractions.

At this point, we have successfully mapped our example system of algebraic data
types into a corresponding system of multivariate generating functions.
Symbolically, our system of multivariate generating functions takes
the general form $\vec{F} = \vec{\Phi}(\vec{F}, \vec{Z})$ where $\vec{F}$ denotes
the vector of generating functions, $\vec{\Phi}$ denotes the vector of corresponding
right-hand side expressions, and $\vec{Z}$ stands for the vector of (all) tuning variables.

First, for $\vec{F} = (L(z,u), D(z,u))$ and $\vec{Z} = (z,u)$ we introduce new
variables, \emph{i.e.}  $\vec{f} = (\lambda, \delta)$ and $\vec{z} = (\zeta, \upsilon)$, respectively.
Next, we apply the following \emph{log-exp
transformation}~\cite{bendkowski_bodini_dovgal_2021} to~\eqref{eq:example-sys}
\begin{equation}
  \vec{F} = \vec{\Phi}(\vec{F}, \vec{Z}) \longrightarrow \vec{f} \geq \log\left(\vec{\Phi}(\exp(\vec{f}), \exp(\vec{z}))\right)
\end{equation}
resulting in
\begin{align}
\delta &\geq \log \left(e^{\zeta} + e^{\zeta + \delta} \right) \nonumber \\
\lambda &\geq \log \left(e^{\delta} + e^{\zeta + 2\lambda} + e^{\zeta + \upsilon + \lambda}\right)
\end{align}
The above two inequalities form \emph{convex} optimisation constraints.
What remains is to formulate the optimisation goal.
In general the optimisation goal takes form
\begin{equation}
  f_{o} - \vec{\mu}^{\intercal} \vec{z} \to \min_{\vec{f}, \vec{z}}
\end{equation}
where $f_{o}$ is the \emph{target type} whose inhabitants we intend to generate,
and $\vec{u}$ is a vector of user-declared expectations matching $\vec{z}$
and thus the introduced tuning parameters.
In our running example the optimisation goal becomes
\begin{equation}
  \lambda - 10,000 \zeta - 2,500 \upsilon \to \min_{\lambda,\delta,\zeta,\upsilon}
\end{equation}
In the end, we constructed a complex optimisation problem from whose solution
we can recover the values for $z, u$ and, $L(z), D(z)$ which realise the
desired Boltzmann model, \emph{cf.}~\cite{bendkowski_bodini_dovgal_2021}.

\subsection{Complexity}
The \emph{parameter tuning} process goes through a few phases, \emph{i.e.} the
problem formulation, running a convex optimisation solver, recovering the
value of the control parameter and respective generating functions, and finally
computing the constructor probabilities for each constructor in the considered
system. The single most expensive phase is finding a proper solution to the
convex optimisation problem.

Luckily, due to the regular shape of algebraic data types, we can leverage
polynomial interior-point algorithms for convex
optimisation~\cite{doi:10.1137/1.9781611970791} and use practically feasible
solvers to achieve parameter tuning. In our current framework, we rely on an
external library called~\texttt{paganini}~\cite{bendkowski_bodini_dovgal_2021}
which allows us to model tuning as a \emph{disciplined convex optimisation
problem} (DCP)~\cite{Grant2006}. The DCP modelling framework can be viewed as a
\emph{domain specific language} which allows its users to systematically build
convex optimisation problems out of simple expressions such as $\log \cdot$ and
$\log \sum \exp^{\cdot}$ though a set of composition rules which follow basic convex
analysis principles. The framework takes care of most tedious tasks such as
formulating the problem in standard form, or providing a feasible starting point
to the solver. While DCP covers a strict subset of the interior-point framework
using so-called conic solvers, problems stemming from algebraic data types
can be effectively expressed and solved.

\begin{figure}[hbt!]
    \begin{center}
        \includegraphics[height=0.02\textheight]{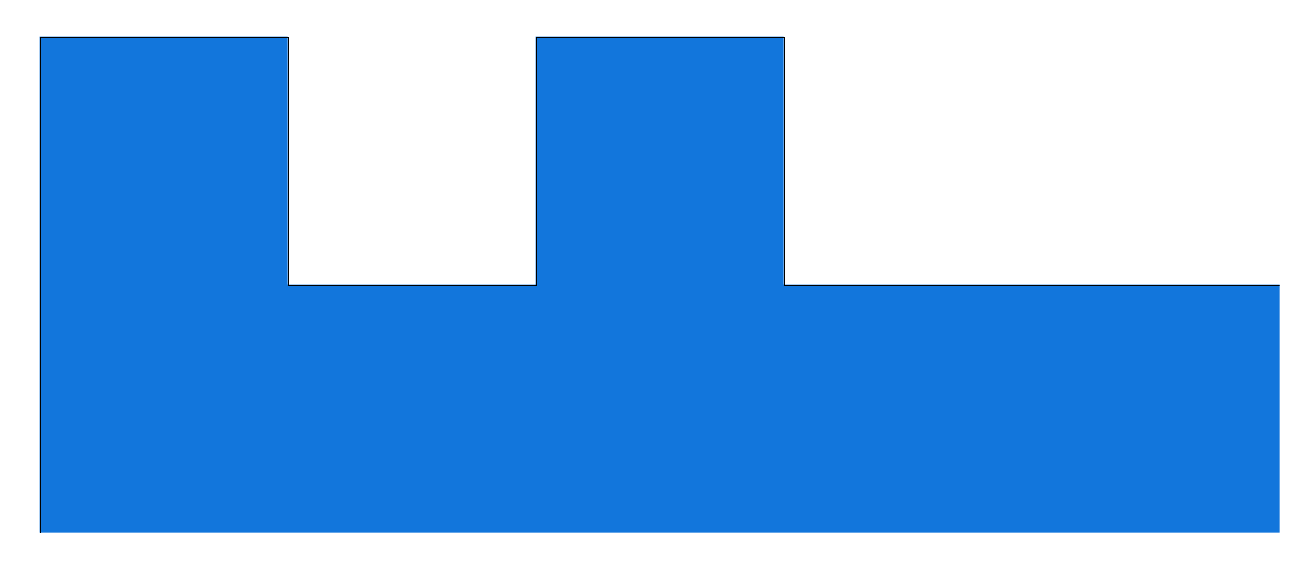} $\ $
        \includegraphics[height=0.02\textheight]{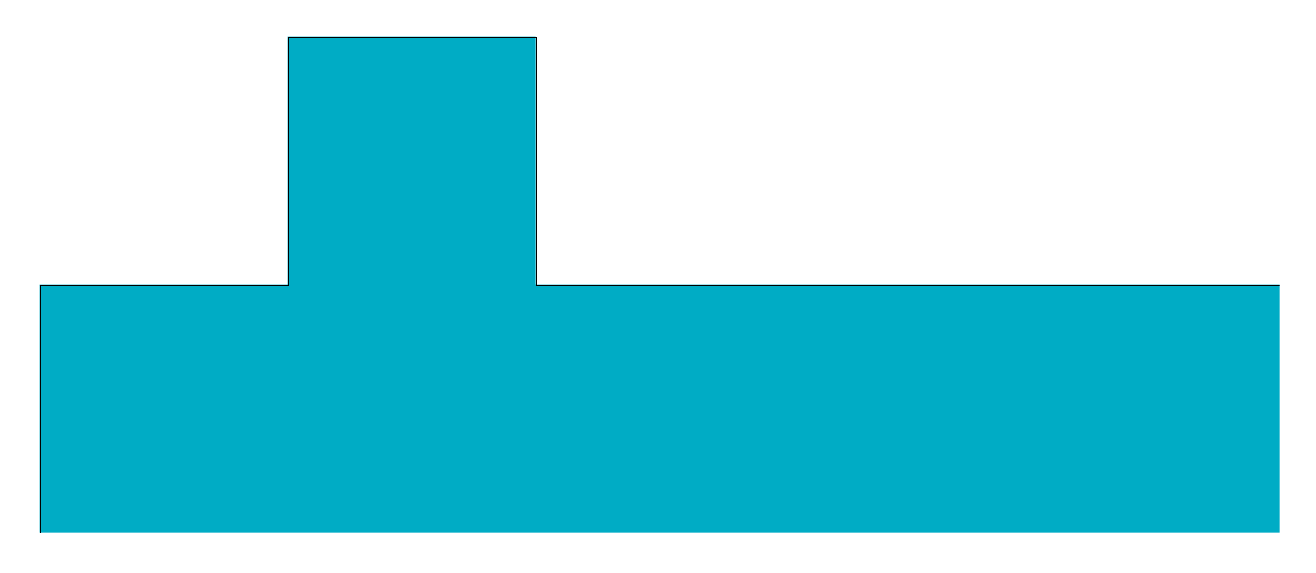} $\ $
        \includegraphics[height=0.02\textheight]{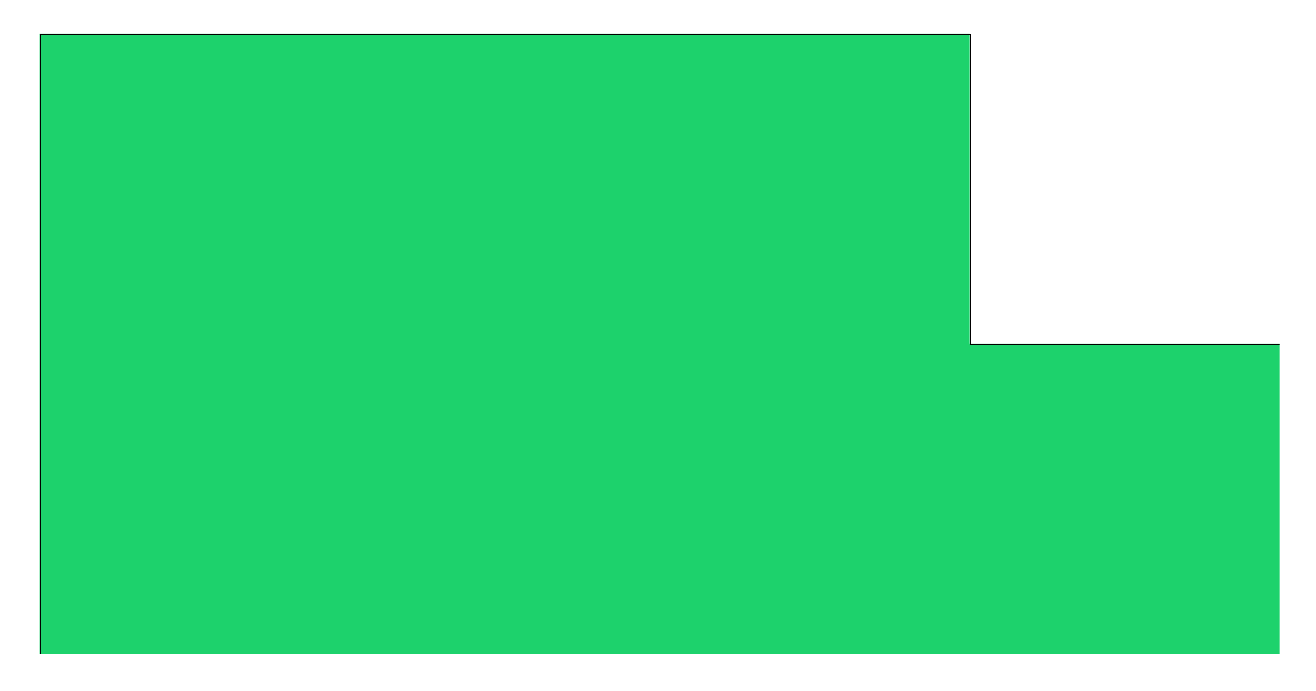} $\ $
        \includegraphics[height=0.02\textheight]{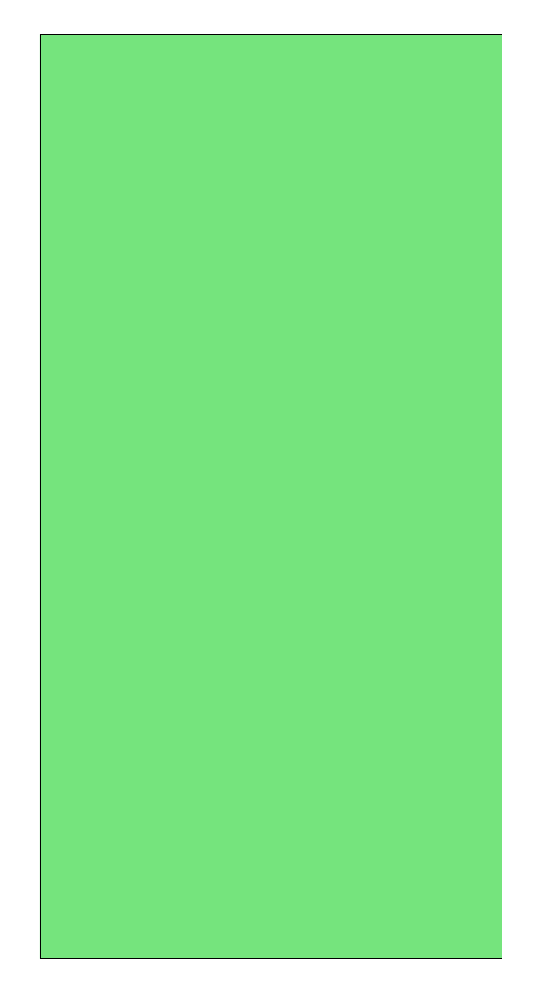} $\ $
        \includegraphics[height=0.02\textheight]{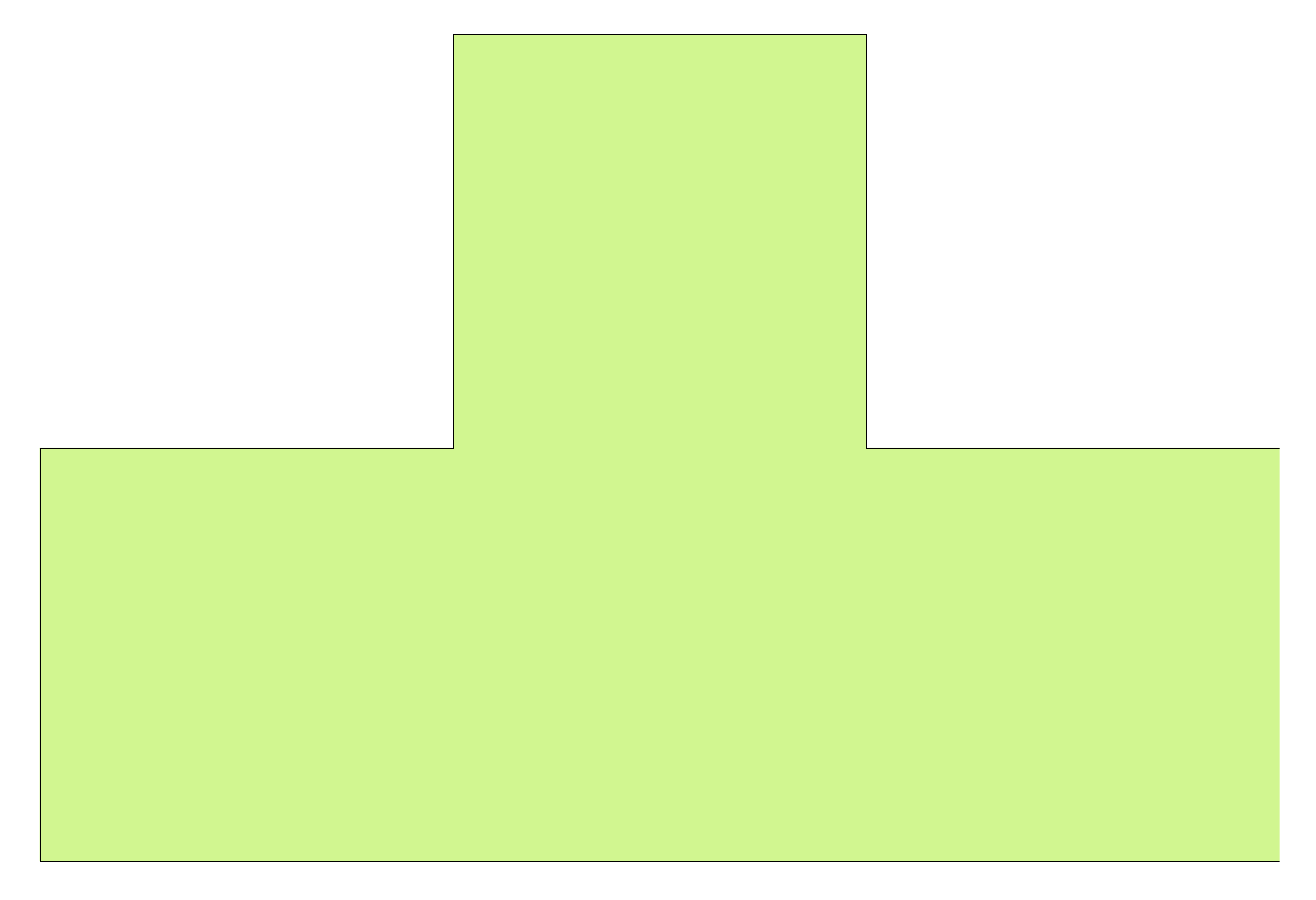} $\ $
        \includegraphics[height=0.02\textheight]{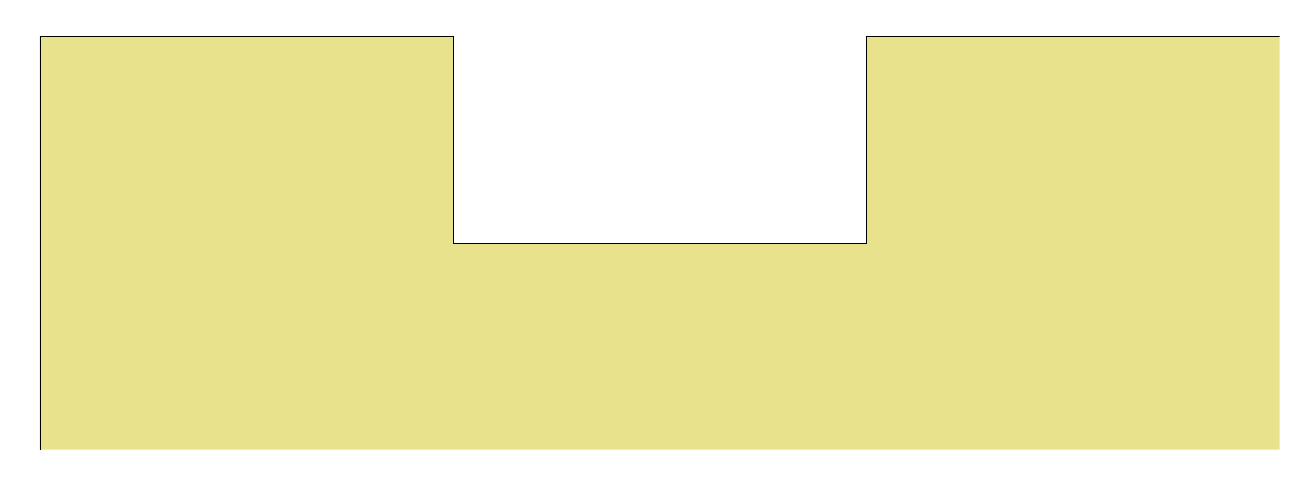} $\ $
        \includegraphics[height=0.02\textheight]{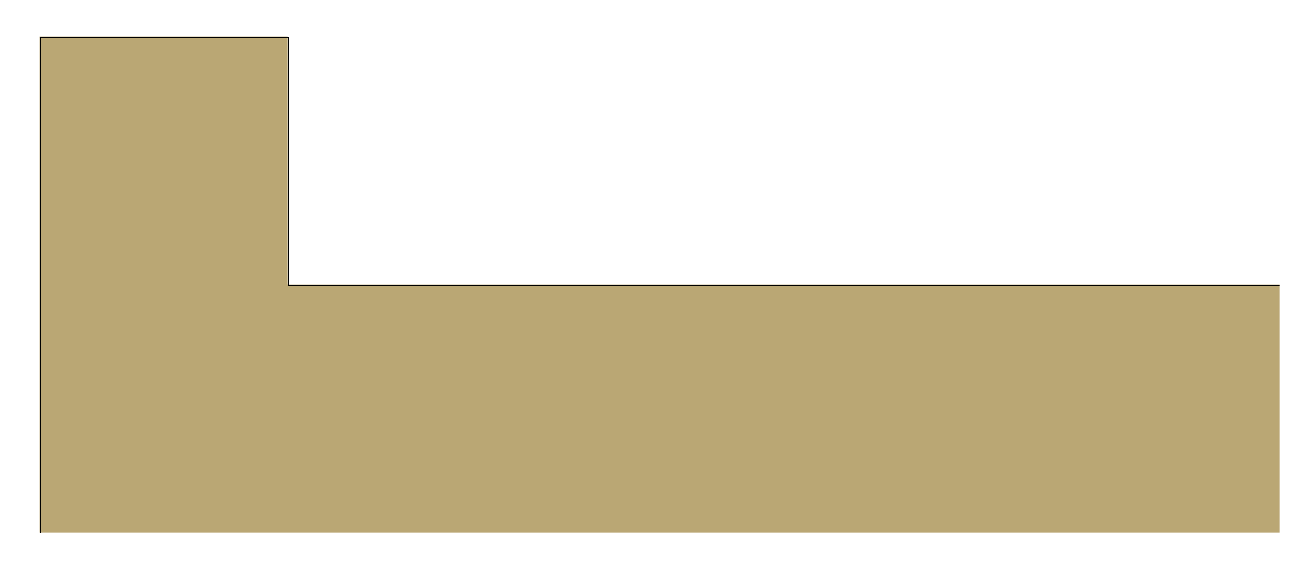} $\ $
        \includegraphics[height=0.02\textheight]{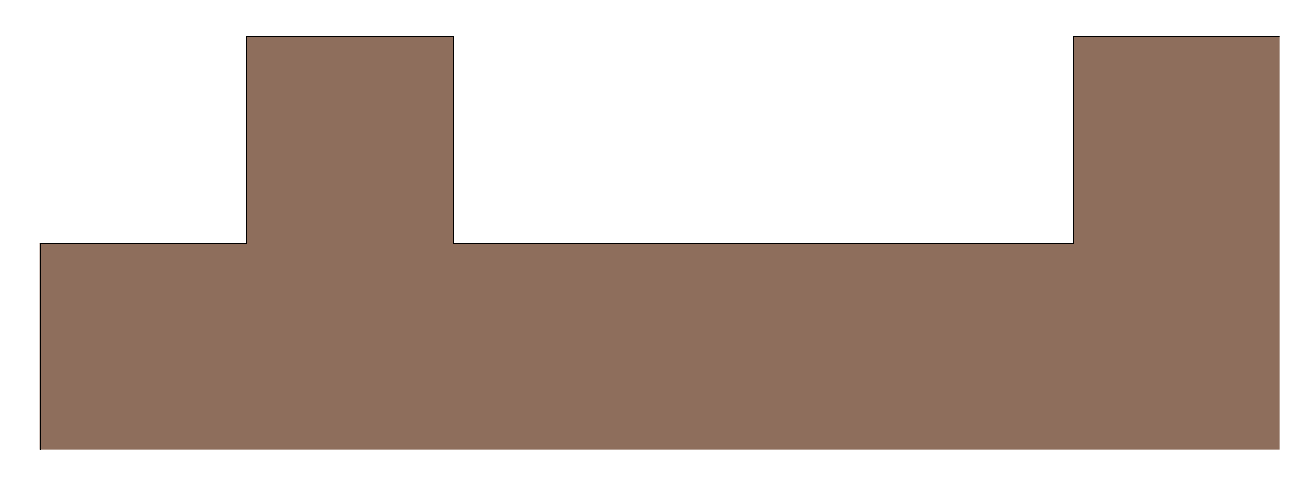} $\ $
        \includegraphics[height=0.02\textheight]{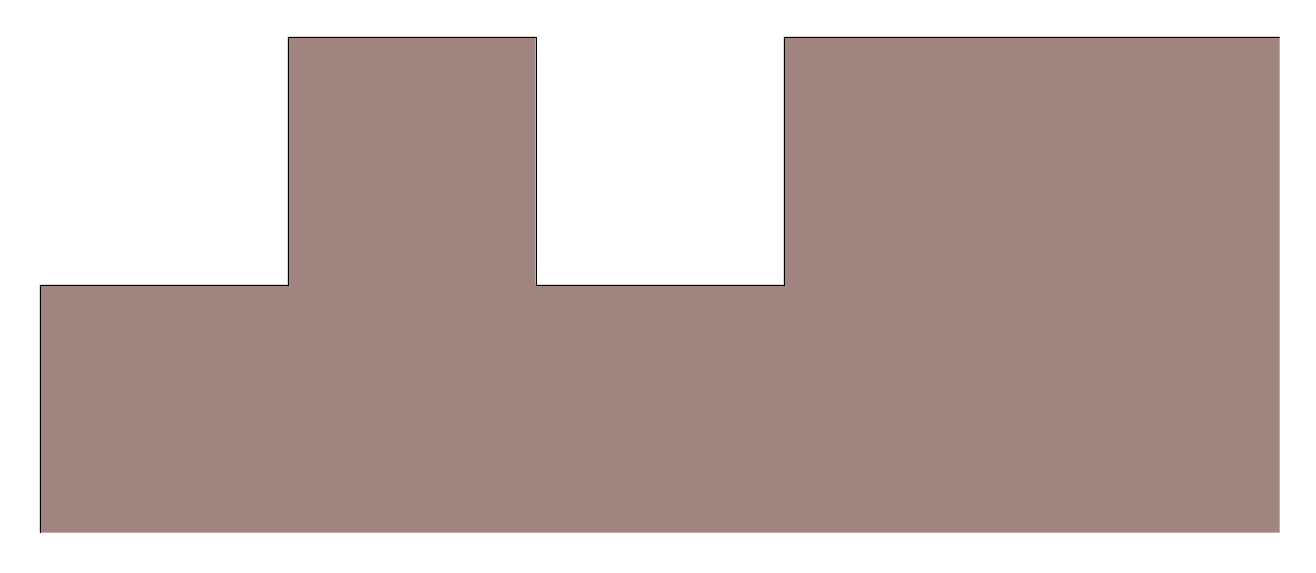} $\ $
        \includegraphics[height=0.02\textheight]{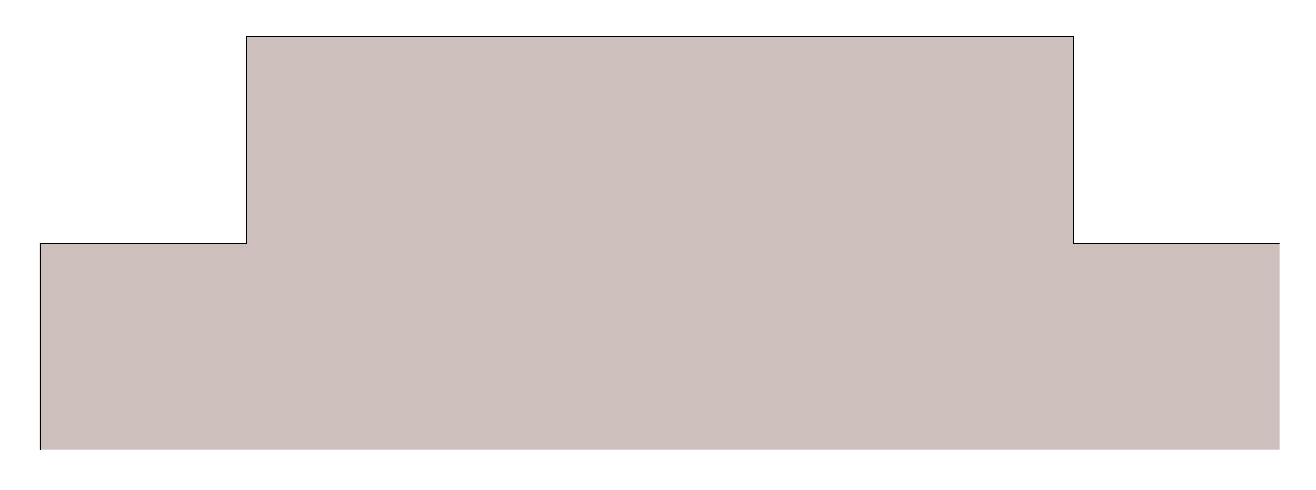}
    \end{center}
    \caption{Examples of admissible tiles.}\label{fig:admissible:tiles}
\end{figure}

\begin{figure}[hbt!]
\centering
  \includegraphics[scale=0.8]{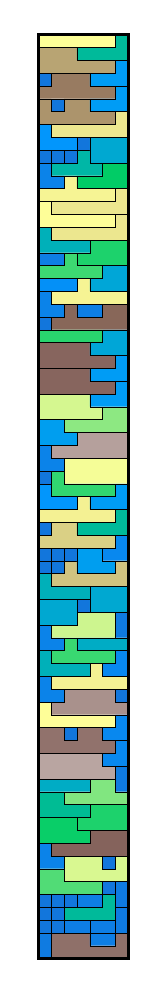}
  \includegraphics[scale=0.8]{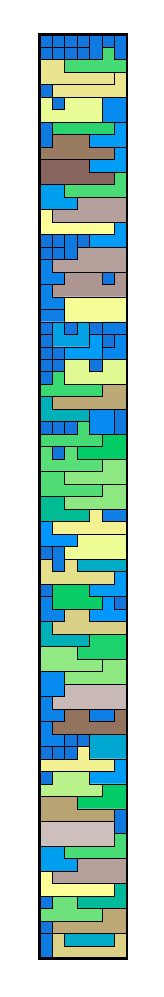}
  \includegraphics[scale=0.8]{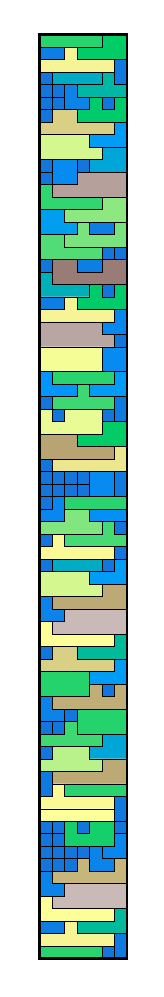}
  \includegraphics[scale=0.8]{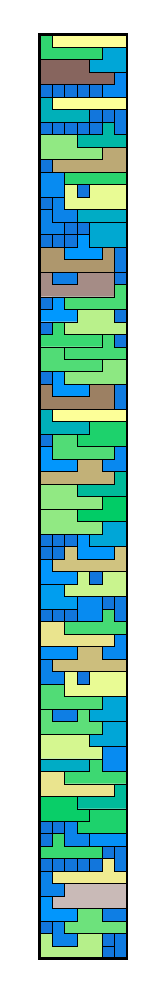}
  \includegraphics[scale=0.8]{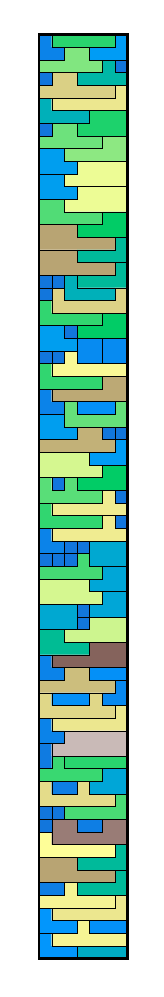}
\caption{Five random ${n \times 7}$ tillings of areas in the interval
$[500; 520]$ using in total $95$ different tiles.}
  \label{fig:tilings-7-126}
\end{figure}

The authors of~\cite{bendkowski_bodini_dovgal_2021} report a benchmark example
of a transfer matrix model tuned using \texttt{paganini}. It consists of a
Boltzmann model generating $n \times 9$ polyomino tillings with over $1,000$
different available tiles. Each tile has a distinct colour and shape,
see~\autoref{fig:admissible:tiles}.
The model was tuned so to achieve outcome polyomino tillings with a
\emph{uniform} colour palette,~\emph{i.e.} each colour occupies in expectation
the same amount of space in each
tilling,~\emph{cf.}~\autoref{fig:tilings-7-126}.  Polyomino tillings of this
form have a corresponding \emph{finite state automaton} with $19,000$ states and
$357,000$ transitions, which was automatically derived as a self-contained Haskell module
used to obtain~\autoref{fig:admissible:tiles}. In our prototype we use the same
\texttt{paganini} library to tune systems of algebraic data types.

Let us remark that we need to approximate the control vector $\hat{x}$ with
precision of order $O(\frac{1}{n})$ to obtain a rejection sampler with linear
time complexity. For a more detailed analysis we invite the curious reader
to~\cite{bendkowski_bodini_dovgal_2021}.

\section{Architecture overview}
With the \emph{tuning as convex optimisation} principle we can tune the
control vector and obtain a Boltzmann model realising the user-declared
values.

Given a system such as
\begin{minted}[fontsize=\small]{haskell}
mkBoltzmannSampler
  System
    { targetType = ''Lambda
    , meanSize = 10_000
    , frequencies = ('Abs, 4_000) <:> def
    , weights =
        ('Index, 0)
          <:> $(mkDefWeights ''Lambda)
    }
\end{minted}
the presented framework must
\begin{itemize}
  \item compute the control vector parameters and related generating function values
        for the user-declared parameter values, and
  \item compile a Boltzmann sampler realising the computed Boltzmann model.
\end{itemize}
These steps involve a series of intermediate transformations which are organised
through a stack of embedded domain specific languages, spread across our Haskell
prototype and the Python tuner \texttt{paganini}. In the following sections we
outline some of the key features of these transformations.

\subsection{Computing constructor distributions}
To compute the required control vector and generating function values, we
convert the system declaration into a \texttt{paganini} input using a custom
monadic eDSL we call
\texttt{paganini-hs}\footnote{\url{https://github.com/maciej-bendkowski/paganini-hs}}.
It is meant as a thin Haskell wrapper around \texttt{paganini}, providing a
convenient and type safe way of expressing the tuning problem.

\begin{minted}[escapeinside=||,mathescape=true,fontsize=\small]{haskell}
formProblem ::
  IO (Either PaganiniError [Maybe Double])
formProblem = paganini' @@ do
  Let z <- variable' 10_000
  Let u <- variable' 4_000

  Let d <- variable
  Let l <- variable

  -- $D(z,u) = z + z D(z,u)$
  d .=. z + z * d

  -- $L(z,u) = D(z, u) + z {L(z, u)}^{2} + z u L(z, u)$
  l .=. d + z * l^2 + z * u * l

  tuneAlgebraic l -- tune for l
  value <$> [z, u, l, d]
\end{minted}
We start with introducing two marking variables
\mintinline[fontsize=\small]{haskell}{z} and
\mintinline[fontsize=\small]{haskell}{u} which correspond to the size of
generated lambda terms and the number of their abstractions, respectively. These
variables are initialised with user-declared values.  Next, we introduce two
more variables \mintinline[fontsize=\small]{haskell}{d} and
\mintinline[fontsize=\small]{haskell}{l} which correspond to
$L(z,u)$ and $D(z,u)$, respectively. Note that these
variables are not \emph{tuned}.

Next, we proceed with mapping type definitions to their corresponding generating
function definitions following the symbolic method outlined
in~\autoref{sec:convex-optimisation}.  For each type, we provide its defining
equation using the \mintinline[fontsize=\small]{haskell}{(.=.)} operator
\begin{minted}[escapeinside=||,mathescape=true,fontsize=\small]{haskell}
(.=.) :: Variable -> Expr -> Spec ()
\end{minted}

Note that on one hand side, we want to conveniently use variables to build more
involved expressions while on the other hand, we do not want expressions to be
used on the left-hand sides of defining equations.  Hence, to lift variables
into expressions we use existential types in the definition of \mintinline[fontsize=\small]{haskell}{Let}:
\begin{minted}[escapeinside=||,mathescape=true,fontsize=\small]{haskell}
newtype Let = Let (forall a . FromVariable a => a)

class FromVariable a where
  fromVariable :: Variable -> a


instance FromVariable Let where
  ...

instance FromVariable Exp where
  ...
\end{minted}
Now, it is possible to safely use variables in the contexts permitting expressions while
keeping \mintinline[fontsize=\small]{haskell}{Variable}s and \mintinline[fontsize=\small]{haskell}{Exp}s
logically separated.

Once the definitions of the type variables are defined, we invoke \texttt{paganini}
using \mintinline[fontsize=\small]{haskell}{tuneAlgebraic l}, and output
the tuned values of $z, u, L(z,u), D(z,u)$.
The whole computation is expressed in a specification monad
\mintinline[fontsize=\small]{haskell}{Spec} which formulates a corresponding
problem in the input format of \texttt{paganini}\footnote{For presentation
purposes we elide boilerplate code handling, e.g.~error handling. The actual
input is slightly more involved.}:
\begin{minted}[escapeinside=||,mathescape=true,fontsize=\small]{python}
import paganini as pg
spec = pg.Specification()
z = pg.Variable(10000)
d = pg.Variable(4000)
l = pg.Variable()
d = pg.Seq(z)

spec.add(l, d + z * u * l + z * l**2)
spec.run_tuner(l, method = pg.Method.STRICT)

print (z.value)
print (u.value)
print (l.value)
print (d.value)
\end{minted}
The \mintinline[fontsize=\small]{haskell}{Spec} monad is a simple state
transformer monad
\begin{minted}[escapeinside=||,mathescape=true,fontsize=\small]{haskell}
type Spec = StateT Program IO
\end{minted}
letting us compose a dedicated \texttt{paganini} \mintinline[fontsize=\small]{haskell}{Program} and
execute it by an external Python interpreter.  Afterwards, the program result,
\emph{i.e.}~the tuning variable values, are collected and returned back as a
Haskell level value.

Let us notice that \texttt{paganini} itself is a \texttt{python} DSL written on
top of \texttt{CVXPY}~\cite{diamond2016cvxpy} --- a modelling library and \emph{de
facto} eDSL for disciplined convex optimisation. It lets us express the tuning
problem in a convenient, domain specific form which can be then transformed into
a readily solvable convex optimisation problem.  Note that in doing so, our
framework does not need to formulate the problem directly, but rather can treat
\texttt{paganini} as a black-box solver.

Finally, let us notice that we retain the original variable names while
composing the \texttt{paganini} program using a source code reification library
\texttt{BinAnn}~\cite{10.1007/978-3-030-57761-2_2}.  Variables names are
reflected in both DSLs.  Such a design choice makes debugging easier and lets
\texttt{paganini-hs} provide better error messages with meaningful variable
names.

\subsection{Sampling from discrete distributions}
Once the values of the control vector and corresponding generating functions are
computed, we can readily calculate the \emph{branching probabilities} for
involved types. Recall that for a type \mintinline[fontsize=\small]{haskell}{a}, the
respective Boltzmann sampler for \mintinline[fontsize=\small]{haskell}{a} has to
make a random decision determining which constructor to use in the process of
generating a random object in \mintinline[fontsize=\small]{haskell}{a}. In our
running example, the \emph{branching probabilities} for type
\mintinline[fontsize=\small]{haskell}{Lambda}
take the form
\begin{equation}
  \frac{D(z,u)}{L(z,u)} \quad
  \frac{z u L(z,u)}{L(z, u)} \quad
  \frac{z {L(z, u)}^{2}}{L(z,u)}
\end{equation}
for \mintinline[fontsize=\small]{haskell}{Index}, \mintinline[fontsize=\small]{haskell}{Abs},
and \mintinline[fontsize=\small]{haskell}{Abs}, respectively, and aptly chosen values of $z$ and $u$.
Hence, in order to choose a constructor
we have to draw a random variable from a \emph{(finite) discrete probability distribution}.
To do so, we can resort to the well-known \emph{inversion method}~\cite{Devr86}; we partition
the interval $[0,1]$ into three segments, each of length corresponding to one of
the available constructors, draw a random real $p$ in between zero and one, and
determine in which segment does $p$ fall into.

While it is possible to choose random constructors using the inversion scheme,
let us remark that it is quite inefficient for our application.  The inversion
method works well under the (unrealistic) real RAM model in which we operate on
real numbers. In practice, we do not have \emph{arbitrary} precision real
numbers, but rather \emph{finite} precision floating-point numbers.  The
inversion scheme samples therefore a random double-precision floating point
number $p \in [0,1]$ to select one of the distribution points. In some cases
the available precision of a single floating-point number might be not enough.
In others, fewer bits are sufficient. For instance,
note that in order to sample from a distribution
$\left(\frac{1}{2},\frac{1}{2}\right)$
a \emph{single} bit is sufficient.

Due to these limitations, we do not use the inversion scheme but rather resort to a
different approach following the \emph{random bit model of sampling} introduced by
Knuth and Yao in~\cite{KY76}.  Instead of using a single floating-point number
to sample from a discrete distribution, one accesses a \emph{lazy} stream of
random bits, consuming one bit at a time. These bits are then used to refine
the search space until a single value can be chosen.

For performance reasons, in our implementation we do not use an actual stream of
bits, but rather use a buffered oracle, as suggested in~\cite{DBLP:journals/corr/abs-1304-1916}.
\begin{minted}[fontsize=\small]{haskell}
data Oracle g = Oracle
  { buffer :: !Word32
  , usedBits :: !Int
  , rng :: g
  }
\end{minted}
The oracle type
\mintinline[fontsize=\small]{haskell}{Oracle g} is parameterised
by a random number generator \mintinline[fontsize=\small]{haskell}{g}.
The oracle consists of a 32-bit buffer and a counter keeping track of how many
random bits have been consumed so far from the current buffer. If the buffer
gets depleted, it can be regenerated as follows
\begin{minted}[fontsize=\small]{haskell}
fresh :: RandomGen g => g -> Oracle g
fresh g = case random g of
    (x, g') -> Oracle { buffer = x
                      , usedBits = 0
                      , rng = g'
                      }
\end{minted}
Using the \mintinline[fontsize=\small]{haskell}{Oracle} type, we can now define
a \mintinline[fontsize=\small]{haskell}{BuffonMachine}\footnote{The name Buffon
  machine was coined by Flajolet, Pelletier and Soria who studied probability distributions which can be simulated \emph{perfectly} using
  a source of unbiased random bits~\cite{DBLP:conf/soda/FlajoletPS11}. While we do not make direct use of their ideas, we consider
  them a source of inspiration for our current work.}
monad for random
computations in the random bit model framework. The
\mintinline[fontsize=\small]{haskell}{BuffonMachine} type is implemented as a
\mintinline[fontsize=\small]{haskell}{newtype} wrapper around the
\mintinline[fontsize=\small]{haskell}{State} monad:
\begin{minted}[fontsize=\small]{haskell}
newtype BuffonMachine g a = MkBuffonMachine
  {runBuffonMachine :: State (Oracle g) a}
  deriving (Functor, Applicative, Monad)
    via State (Oracle g)
\end{minted}

Using the ideas of~\cite{KY76} it is possible to construct an entropy-optimal
\emph{discrete distribution-generating tree} (DDG) implementing a sampler for
any discrete distribution $P$ of rational numbers.  In other words, it is
possible to construct a sampler for $P$ which uses the least average number of
random bits to sample from $P$. Unfortunately, the entropy-optimal DDGs can be
\emph{exponentially} large in the number of bits required to encode the input
distribution $P$. For instance, the binomial distribution $\mathbb{B}(n, p)$
with parameters $n = 50$ and $p = \frac{61}{500}$ requires a DDG of height
$10^{104}$, see~\cite{10.1145/3371104}.

Unfortunately, such an overhead renders DDGs virtually impractical.  Due
to that, we use recently developed \emph{approximate sampling
schemes}~\cite{10.1145/3371104} which are a practical trade-off between the
entropy perfect DDGs and feasible, finite precision sampling algorithms.
Instead of sampling from a discrete probability distribution
$P = (p_{1},\ldots,p_{n})$ we find an entropy optimal sampling algorithm for a
closest approximation $\hat{P} = (\hat{p}_{1},\ldots,\hat{p}_{n})$ of $P$ among all
sampling algorithms which operate within a finite $k$-bit precision. Let us note that
the framework of approximate sampling schemes, and in particular its prototype implementation\footnote{
  https://github.com/probcomp/optimal-approximate-sampling}, supports
several statistical measures of \emph{approximation error} between
probability distributions, including Kullback-Leibler,
Pearson chi-square, and Hellinger divergence.

The optimal approximate distribution $\hat{P}$ can be readily found as soon as the
constructor distribution is computed in so-called linear, compact vector form.  We use
a prototype implementation of optimal approximate sampling algorithms to find the
compact vector form of DDGs. Compiled Boltzmann samplers readily choose
constructors from the compact DDGs represented as
\mintinline[fontsize=\small]{haskell}{Vector Int}.

\subsection{Anticipated rejection}
A straightforward implementation of Boltzmann samplers
\begin{minted}[fontsize=\small]{haskell}
  sample :: RandomGen g => BuffonMachine g a
\end{minted}
has some practical drawbacks. While the underlying Boltzmann model provides
control over the \emph{mean size} of its outcomes, we have no finer control over
the \emph{actual} size of generated objects.  In some cases, the outcome sample
size might be significantly larger than the user-declared mean size. Without any
additional control, Boltzmann samplers might consume significantly more
resources than required.

In the presented framework we implement Boltzmann samplers with
\emph{anticipated rejection}, see~\cite{BodGenRo2015}. The idea is quite
simple. The user provides an \emph{upper bound} on the size of generated
outcomes\footnote{Note that this is also the recommended generator design choice of
QuickCheck.}. During generation we maintain the current size of the sample. If it
exceeds the given upper bound, the process is terminated and the sample is
\emph{rejected}. Consequently, the signature of
\mintinline[fontsize=\small]{haskell}{sample} becomes
\begin{minted}[fontsize=\small]{haskell}
  sample :: RandomGen g =>
            UpperBound ->
            MaybeT (BuffonMachine g) (a, Int)
\end{minted}
To give the user a more fine-grained control over the outcome size of
sampled objects, the user can provide an \emph{admissible} size range
$[(1-\varepsilon)n, (1+\varepsilon)]$. The framework samples objects until one with
admissible size is generated. Note that such a \emph{rejection} scheme guarantees
that inadmissible samples are rejected as soon as possible.\footnote{The
  same idea can be readily applied to \emph{all} parameters. Our prototype
  framework supports only anticipated rejection for the outcome size.}
\begin{minted}[fontsize=\small]{haskell}
toleranceRejectionSampler ::
  (RandomGen g, BoltzmannSampler a) =>
  MeanSize -> Double -> BuffonMachine g a
toleranceRejectionSampler n eps =
  rejectionSampler lb ub
  where
    lb = MkLowerBound $
      floor $ (1 - eps) * fromIntegral n
    ub = MkUpperBound $
      ceiling $ (1 + eps) * fromIntegral n
\end{minted}

Let us remark that anticipated rejection has an tiny impact on the underlying
Boltzmann model. As we limit admissible sizes, we also impose a small bias
in the distribution, initially not taken into account. Indeed, objects of
inadmissible sizes can no longer be sampled, and so their total probability mass
gets redistributed among admissible objects.  In consequence, the
original tuning goal $\mathbb{E}(N) = n$ must be modified to accommodate an
additional \emph{bias parameter} $\delta$ such that $\mathbb{E}(N) = \delta n$,
\emph{cf}.~\cite{bendkowski_bodini_dovgal_2021}. The specific value of $\delta$
depends on the type of parameter corresponding to the variable $N$ and, in
particular, its corresponding asymptotic behaviour in the related system of
multivariate generating functions. While our prototype implementation does
not introduce the bias parameter, let us remark that it can diminish the
overall number of rejections required to find an admissible sample.
For more details we invite the curious reader
to~\cite{bendkowski_bodini_dovgal_2021}.

\subsection{ADT and newtype samplers}
Boltzmann samplers for algebraic data structures have a regular format.  For
instance, our running example of \mintinline[fontsize=\small]{haskell}{Lambda}
has the following\footnote{For the reader's convenience, we elide boilerplate code which
clouds the structure of the algorithm.} Boltzmann sampler:
\begin{minted}[fontsize=\small]{haskell}
instance BoltzmannSampler Lambda where
  sample ub =
    do guard (ub >= 0)
      lift (BuffonMachine.choice ...)
          >>=
            (\case
              0 -> do (x_0, w_0) <- sample ub
                      pure (Index x_0, w_0)
              1 -> do (x_0, w_0) <- sample (ub - 1)
                      (x_1, w_0) <- sample (ub - w_0 - 1)
                      pure (App x_0 x_1, 1 + w_0 + w_1)
              2 -> do (x_0, w_0) <- sample (ub - 1)
                      pure (Abs x_0, 1 + w_0)
  \end{minted}
The \mintinline[fontsize=\small]{haskell}{sample} function has a single
parameter \mintinline[fontsize=\small]{haskell}{ub} which defines a \emph{size
budget} which the sampler cannot overreach, as enforced by
\mintinline[fontsize=\small]{haskell}{guard (ub >= 0)}.
If the sampler has some non-negative size budget left, it can proceed with
generating the object. To do so, the sampler draws a random
number according to the respective constructor distribution.
The choice function has signature
\begin{minted}[fontsize=\small]{haskell}
choice :: RandomGen g => Distribution -> Discrete g
  \end{minted}
where
\begin{minted}[fontsize=\small]{haskell}
newtype Distribution =
  MkDistribution {unDistribution :: Vector Int}
  deriving stock (Show)

type Discrete g = BuffonMachine g Int
  \end{minted}
represent the compact linear DDG, and discrete random integer variables.
Note that the actual distribution is inserted directly in the body of
the sampler function.

Next, the generated random number is mapped onto a concrete constructor.  We use
\mintinline[fontsize=\small]{haskell}{sample} to generate all of the constructor
parameters. At the same time, we keep track of the size budget accounting for
the weight of the considered constructor and size of each generated
subexpression.

Such a Boltzmann sampler construction easily generalises onto arbitrary
algebraic data types. However, since samplers are implemented as instances of
the \mintinline[fontsize=\small]{haskell}{BoltzmannSampler} type class, we can
have at most one sampler for each type. In some circumstances, we might want to
have multiple samplers for the same type. To support such use cases, we support
the compilation of Boltzmann samplers for
\mintinline[fontsize=\small]{haskell}{newtype} synonyms.  Note that the
structure of such Boltzmann samplers is \emph{almost} the same as for regular
data types. To support them, we need to change the constructor distribution, and
adjust the return type of generated object.  The former can be achieved through
a separate tuning problem. The latter, on the other hand, through safe,
zero-cost constructor type coercions~\cite{10.1145/2628136.2628141}. For each
constructor application we introduce an explicit coercion which changes the
constructor type so match the \mintinline[fontsize=\small]{haskell}{newtype}
synonym.  For instance, for $\lambda$-term application we use
\begin{minted}[fontsize=\small]{haskell}
  coerce
    @(Lambda -> Lambda -> Lambda)
    @(BinLambda -> BinLambda -> BinLambda) App
  \end{minted}
  instead of \mintinline[fontsize=\small]{haskell}{App}. Note
  that such a coercion imposes correct type constraints on the
  argument \mintinline[fontsize=\small]{haskell}{sample} corresponding to the considered constructor.

\subsection{Generated class instances}
In principle, for a given system of algebraic data types there exists a variety
of different Boltzmann samplers differing only in the respective branching
probabilities.  It is the specific tuning parameters which determine the
\emph{exact} dependencies among the Boltzmann samplers corresponding to the
system's types.  For instance, in our running example
\begin{minted}[fontsize=\small]{haskell}
data DeBruijn
  = Z
  | S DeBruijn

data Lambda
  = Index DeBruijn
  | App Lambda Lambda
  | Abs Lambda
\end{minted}
the data type \mintinline[fontsize=\small]{haskell}{Lambda} depends on
\mintinline[fontsize=\small]{haskell}{DeBruijn}. Consequently, when we generate
an instance of \mintinline[fontsize=\small]{haskell}{BoltzmannSampler Lambda} we
need to know the correct \mintinline[fontsize=\small]{haskell}{sample} instance
to invoke in order to sample DeBruijn indices.

For regular \mintinline[fontsize=\small]{haskell}{data} types, we simultaneously derive
separate \mintinline[fontsize=\small]{haskell}{BoltzmannSampler} class instances
for each type in the system. In consequence, users can access Boltzmann samplers
of all of the system types, not just the target one. Alas, it also means that
it is not possible to support multiple samplers for the \emph{exact same} data type
as we would have clashing instances of \mintinline[fontsize=\small]{haskell}{BoltzmannSampler}.
In order to enable multiple samplers for the same type, we assume a different
derivation strategy for target \mintinline[fontsize=\small]{haskell}{newtype}s.
Let us again recall the example of binary lambda terms:
\begin{minted}[fontsize=\small]{haskell}
newtype BinLambda = MkBinLambda Lambda
\end{minted}
Since generating \mintinline[fontsize=\small]{haskell}{BoltzmannSampler}
instances for the underlying types \mintinline[fontsize=\small]{haskell}{Lambda}
and \mintinline[fontsize=\small]{haskell}{DeBruijn} would lead to ambiguous
class instances, we instead generate a
\mintinline[fontsize=\small]{haskell}{newtype} synonym for
\mintinline[fontsize=\small]{haskell}{DeBruijn}
\begin{minted}[fontsize=\small]{haskell}
newtype Gen_DeBruijn = MkGen_DeBruijn DeBruijn
\end{minted}
where \mintinline[fontsize=\small]{haskell}{Gen_DeBruijn} is a fresh, unique
type name. Next, we derive
\mintinline[fontsize=\small]{haskell}{BoltzmannSampler} instances for both
\mintinline[fontsize=\small]{haskell}{BinLambda} and
\mintinline[fontsize=\small]{haskell}{Gen_DeBruijn}. Whenever
a sampler for \mintinline[fontsize=\small]{haskell}{DeBruijn} is required,
we use a type coercion to the associated \mintinline[fontsize=\small]{haskell}{Gen_DeBruijn}, instead.
For instance, in our running example we coerce \mintinline[fontsize=\small]{haskell}{Index} as follows
\begin{minted}[fontsize=\small]{haskell}
  coerce
    @(DeBruijn -> BinLambda)
    @(Gen_DeBruijn -> BinLambda) Index
  \end{minted}
Because the generated \mintinline[fontsize=\small]{haskell}{newtype}s are unique,
it is possible to declare multiple Boltzmann samplers for the same underlying
type without ambiguous class instances.

Let us remark that the presented derivation strategy carries an important
advantage. Specifically, in the presented approach it is possible to access
Boltzmann samplers for all types in the tuned system, which might be especially
important if the system has multiple types users need to sample from
(\emph{e.g.}~when the system represents a context-free grammar).  Unfortunately,
such a design decision also forces users to create
\mintinline[fontsize=\small]{haskell}{newtype}s if multiple samplers for the
same type are required. We could avoid
\mintinline[fontsize=\small]{haskell}{newtype} wrappers using auto-generated
anonymous sampler functions for each system, however then users would have no
direct and convenient access to these sampler functions.

\subsection{Known limitations}
Multi-parametric Boltzmann samplers support systems of (possibly mutually
recursive) \emph{non-parametric} algebraic data types,~\emph{i.e.} ADTs of kind
\mintinline[fontsize=\small]{haskell}{*}. Parametric ADTs, such as
\begin{minted}[fontsize=\small]{haskell}
data BinTree a
  = Node (BinTree a) (BinTree a)
  | Leaf a
\end{minted}
of kind \mintinline[fontsize=\small]{haskell}{(* -> *)} do not have a
corresponding Boltzmann model as, \emph{a priori}, the structure and size of
objects of type \mintinline[fontsize=\small]{haskell}{a} are unknown.  Depending
on the concrete instantiation of \mintinline[fontsize=\small]{haskell}{a}, the
constructor distribution for \mintinline[fontsize=\small]{haskell}{BinTree} can
vary. While it is possible to define Boltzmann models for
\mintinline[fontsize=\small]{haskell}{BinTree a} where
\mintinline[fontsize=\small]{haskell}{a} are concrete, non-parametric types, our
current prototype implementation does not support this.

Moreover, we deliberately do not provide default Boltzmann samplers for certain
primitive types, such as \mintinline[fontsize=\small]{haskell}{Bool} or
\mintinline[fontsize=\small]{haskell}{Integer}. The former is a type with
finitely many inhabitants and thus Boltzmann models should not be
preferred\footnote{Let us notice that it \emph{is} possible to define Boltzmann
models for finite types, but other, more direct and simple sampling methods are
available.}. While the latter is an \emph{infinite type}, there is no universal
or default size notion attached to integers. In certain contexts a unary
encoding of integers might be used, as for instance in the case of $\lambda$-terms in
the DeBruijn notation, whereas in others a compact binary one might be more
appropriate. We choose not impose a default size notion and leave the decision
to the user.

Let us also remark that for certain size notions or requested
constructor frequencies there might be no corresponding Boltzmann model.
For instance, consider
\begin{minted}[fontsize=\small]{haskell}
data BinTree
  = Node BinTree BinTree
  | Leaf
\end{minted}
where \mintinline[fontsize=\small]{haskell}{Node} contributes weight one and
\mintinline[fontsize=\small]{haskell}{Leaf} contributes no weight at all.  In
other words, the number of \mintinline[fontsize=\small]{haskell}{BinTree}s of
size $n$ corresponds to the $n$th Catalan number.  While
\mintinline[fontsize=\small]{haskell}{BinTree} has a well-defined Boltzmann
model under the assumed size notion,
\mintinline[fontsize=\small]{haskell}{[BinTree]} does not. Note that there is an
infinite number of lists of \mintinline[fontsize=\small]{haskell}{BinTree}s of
size zero
\begin{minted}[escapeinside=||,mathescape=true,fontsize=\small]{haskell}
[Leaf], [Leaf, Leaf], [Leaf, Leaf, Leaf], |$\ldots$|
\end{minted}
There exists therefore no \emph{uniform} distribution of
\mintinline[fontsize=\small]{haskell}{BinTree} lists of size $n$ and so, there
is no corresponding Boltzmann model for
\mintinline[fontsize=\small]{haskell}{[BinTree]}. Such systems are called
\emph{ill-founded} and can, in principle, be recognised before the tuning
procedure is initiated, \emph{cf.}~\cite{pivoteau2012}. Let us remark that in our prototype
implementation we do not implement well-foundness checks, and
hence let the compilation fail when the tuner is invoked.

\section{Benchmarks}
To benchmark the run-time performance of our prototype implementation
we use the following example system of $\lambda$-terms in DeBruijn notation:
\begin{minted}[fontsize=\small]{haskell}
data DeBruijn
  = Z
  | S DeBruijn

data Lambda
  = Index DeBruijn
  | App Lambda Lambda
  | Abs Lambda

mkBoltzmannSampler
  System
    { targetType = ''Lambda
    , meanSize = 1000
    , frequencies = def
    , weights =
        ('Index, 0)
          <:> $(mkDefWeights ''Lambda)
    }

lambdaSampler :: Int -> IO [Lambda]
lambdaSampler n =
  evalIO $
    replicateM n $
      rejectionSampler @SMGen
        (MkLowerBound 800) (MkUpperBound 1200)
\end{minted}
We request a (univariate) Boltzmann sampler tuned so to generate
random $\lambda$-terms with expected size $1,000$. We measure the performance
of a rejection sampler generating $\lambda$-terms of sizes in between $800$
and $1,200$. In other words, we tolarate a $20\%$ size deviation from
the expected target size.

We present three sets of
criterion\footnote{\url{https://hackage.haskell.org/package/criterion}}
benchmarking suites, generating 10, 100, and 1,000
random samples:
\begin{center}
\begin{tabular}{ c | c }
 mean time & 10.95 ms \\
  \hline
 standard deviation & 882.9 $\mu$s
\end{tabular}
\end{center}

\begin{center}
\begin{tabular}{ c | c }
 mean time & 104.8 ms \\
  \hline
 standard deviation & 5.67 ms
\end{tabular}
\end{center}

\begin{center}
\begin{tabular}{ c | c }
 mean time & 1.127 s \\
  \hline
 standard deviation & 40.73 ms
\end{tabular}
\end{center}
Note that generating a single $\lambda$-term of size in between $800$ and $1,200$
takes, on average, around $1.1$ ms.

An analogous sampler generating $100$ samples of target mean size $10,000$ and a
smaller $10\%$ tolerance has a similar performance:
\begin{center}
\begin{tabular}{ c | c }
 mean time & 3.064 s \\
  \hline
 standard deviation & 119.5 ms
\end{tabular}
\end{center}
In this case, generating a single $\lambda$-term of size in between $9,000$
and $11,000$ takes, on average, $30$ ms.

Generating $100$ terms of even larger mean size $100,000$ and the same, $10\%$
tolerance gives the following benchmark:
\begin{center}
\begin{tabular}{ c | c }
 mean time & 26.16 s \\
  \hline
 standard deviation & 1.371 s
\end{tabular}
\end{center}
Note that the sampler performance scales linearly with the target mean size.
Generating a single $\lambda$-term of size in between $90,000$
and $110,000$ takes, on average, $260$ ms.

Finally, we present a benchmark example generating $10$ random $\lambda$-terms of mean
size $1,000,000$ and a $10\%$ size tolerance:
\begin{center}
\begin{tabular}{ c | c }
 mean time & 42.07 s \\
  \hline
 standard deviation & 8.128 s
\end{tabular}
\end{center}
Note that, on average, sampling a random $\lambda$-term takes just $4.2$ s.
It is therefore feasible to generate even larger $\lambda$-terms.

\section{Related work}

\subsubsection*{Boltzmann samplers}
Automatic compilation of Boltzmann samplers for algebraic data types was first
implemented in Objective Caml~\cite{10.1145/1596627.1596637}. While similar in
spirit to the presented work, compiled samplers do not support multi-parametric
tuning.  Branching probabilities are computed using a combinatorial Newton
method developed in~\cite{pivoteau2012}.

A similar \texttt{boltzmann-samplers} framework for the automatic compilation of
Boltzmann samplers was developed for
Haskell\footnote{\url{https://hackage.haskell.org/package/boltzmann-samplers}}.
This framework, however, does not support multi-parametric tuning. It uses
similar ideas to~\cite{10.1145/1596627.1596637} including the idea of
\emph{pointed-specifications} and \emph{singular samplers}\footnote{
Let us note that it is possible to approximate them with arbitrary precision using
large, finite parameter values.} with infinite mean target
size~\emph{cf.}~\cite{BodGenRo2015}. Constructors are sampled using the
inversion method. To compare \texttt{boltzmann-samplers} with our prototype
framework, we used a rejection-based sampler to generate $100$ random $\lambda$-terms
of sizes in between $800$ and $1,200$ (using $1,000$ as the expected target
size).
\begin{center}
\begin{tabular}{ c | c }
 mean time & 9.715 s \\
  \hline
 standard deviation & 671.5 ms
\end{tabular}
\end{center}
Note that \texttt{boltzmann-samplers} is over $92$ times slower than an analogous
sampler compiled using our framework. A ceiled, rejection-based singular sampler
performs a bit better, although it is still over $54$ times slower
\begin{center}
\begin{tabular}{ c | c }
 mean time & 5.689 s \\
  \hline
 standard deviation & 224.3 ms
\end{tabular}
\end{center}

\subsubsection*{Branching processes}
In QuickCheck~\cite{Claessen-2000}, the prominent framework for random testing
in Haskell, users can control the outcome distribution of user-declared
generators through, among other things, custom constructor weights influencing
the constructor distribution. Unfortunately, it is quite difficult to
\emph{rigorously} control the outcome distribution of so-defined generators.  To
overcome these challenges, the authors of~\cite{10.1145/3242744.3242747} proposed
to adopt \emph{branching processes} to derive QuickCheck generators.

In this approach, the user-declared target outcome distribution is used to
compute an apt map of constructor weights leading to QuickCheck generators satisfying the
requested constructor distribution. These computations are performed at
compile-time and so, similarly to Boltzmann model tuning, there is no
additional run-time overhead.

As with other frameworks, we compared our prototype implementation with \texttt{DRaGeN}
implementing the ideas of~\cite{10.1145/3242744.3242747} using our running example of
generating $\lambda$-terms of mean size $1,000$ and a uniform outcome distribution:
\begin{center}
\begin{tabular}{ c | c }
 mean time & 366.6 ms \\
  \hline
 standard deviation & 32.46 ms
\end{tabular}
\end{center}
Note that in this benchmark, our prototype is more than $3$ times faster.

Let us remark that there is a significant difference in compilation times
between \texttt{DRaGeN} and our prototype. The branching process computations
require time which is proportional to the target size, unlike Boltzmann model
tuning which depends on the \emph{bit representation length} of this value.
Consequently, Boltzmann samplers can be compiled much quicker allowing users
to derive samplers for significantly larger mean parameter values.

\subsubsection*{Enumeration generators}
If uniform outcome distribution is required, one can resort to \emph{enumerative}
random generators which injectively encode the inhabitants of a target algebraic data type
to consecutive natural numbers. In addition, such maps are size-monotonic and
so inhabitants of equal size correspond to a range of natural numbers $[n_{1},n_{2}]$.
It is therefore possible to leverage a natural number generator to uniformly sample from
$[n_{1},n_{2}]$ and \emph{decode} a corresponding inhabitant by inverting the
encoding map.

The \texttt{feat}~\cite{10.1145/2430532.2364515} library is one prominent
example of such a sampling scheme in Haskell.  It supports the enumeration of
algebraic data types, and (uniform) random generation of their inhabitants.
Generating $100$ random $\lambda$-terms of (exact) size $1,000$ has the following
performance
\begin{center}
\begin{tabular}{ c | c }
 mean time &  174.1 ms \\
  \hline
 standard deviation & 16.45 ms
\end{tabular}
\end{center}
Note that \texttt{feat} is more that $1.6$ times slower than Boltzmann sampler
with mean size $1,000$ and a $10\%$ size tolerance. Our sampler outperforms the
\texttt{feat} one even in the case of a $1\%$ size tolerance. In the
\emph{exact-size} sampling regime, however, implemented Boltzmann samplers are
no longer linear, but have a quadratic $O(n^{2})$ average runtime complexity.
Consequently, for small or moderate sizes where enumeration generators are
feasible \texttt{feat} becomes more efficient.

Let us remark, however, that recent theoretical improvements to Boltzmann
samplers in the exact-size regime have brought their $O(n^{2})$ average
complexity down to $O(n)$,
\emph{cf.}~\cite{https://doi.org/10.48550/arxiv.2105.12881,https://doi.org/10.48550/arxiv.2110.11472}.

\section{Conclusions}
We presented a novel framework for the automatic derivation of multi-parametric
Boltzmann samplers. With a clean separation of concerns, we provided a
declarative and highly modular prototype Haskell implementation which matches,
or vastly outperforms several prominent random generation frameworks in a
moderate to large outcome size regime.

Given a set of user-declared size notion, target constructor frequencies and
size, our framework synthesises efficient, entropy-optimal Boltzmann samplers.
Suitable branching probabilities are obtained at compile time through a series
of conceptually simpler, intermediate steps. First, the tuning problem is
expressed in a specialised and type safe eDSL called \texttt{paganini-hs}. The
eDSL composes an optimisation problem in another, python-based DSL called
\texttt{paganini}.  There, the domain optimisation problem is further broken
down into a convex optimisation problem expressed in yet another DSL called
\texttt{CVXPY}. The \texttt{CVXPY} framework chooses a suitable solver, finds an
apt starting point, and solves the convex optimisation problem. Its result is
then hoisted through the series of eDSL back into to our framework.  Let us
notice that each of these intermediate steps forms a separate conceptual module
in our framework, each having a clean, distinct set of responsibilities. In
particular, the tuning engine forms a separate module which can be used for other
purposes or in other frameworks.

While our framework is not unduly optimised, it already exhibits the practical
potential of Boltzmann samplers in the field of random generation, so
prominently used throughout the functional programming language community. Our
benchmarks suggest that Boltzmann samplers are an effective tool in generating
large, random inhabitants of algebraic data types. With the additional feature
of parameter tuning, it is possible to control not only the size of generated
objects, but also their expected shape and form, such as the
constructor frequencies. Consequently, multi-parametric Boltzmann samplers
form a versatile random generation platform combining rigorous control
over the outcome distribution with a convenient, declarative user interface.

\begin{acks}
  We would like to express our thanks to Olivier Bodini, Matthieu Dien, Sergey
Dovgal, Agustín Mista, Pierre Lescanne, Martin Pépin, and Li-yao Xia for our
encouraging discussions. This work was partially supported by the ANR project
LambdaComb (ANR-21-CE48-0017).
\end{acks}

\bibliographystyle{ACM-Reference-Format}
\bibliography{references}

\end{document}